\documentclass[12pt]{iopart}

\newcommand{\be}{\begin{eqnarray}}
\newcommand{\ee}{\end{eqnarray}}
\newcommand{\rar}{\rightarrow}

\begin{document}

\title[
Massive black holes and strange stars
] 
{New data on young and old black holes and other unexpected creatures
}

\author{A.D. Dolgov
}

\address{
 Novosibirsk State University, Novosibirsk, Russia \\
 ITEP, Moscow, Russia

}
\ead{dolgov@fe.infn.it
}
\vspace{10pt}
\begin{indented}
\item[]August 2019
\end{indented}

\begin{abstract}

A review on recent astronomical observations indicating to unexpectedly abundant population of the 
contemporary and $ z \sim 10 $ universe  by massive black holes in all mass ranges are is presented. 
It is argued that these black holes are mostly primordial. The data on some other stellar-kind objects 
which are also may be primordial are discussed.

\end{abstract}

%
%
%
%
%

\section{Introduction \label{s-intro}}

{Recent astronomical data, which keep on appearing almost every day, show that the contemporary, {${z \sim 0}$,}  
and early,  {${z \sim 10}$,}  universe is much more abundantly populated by all kind of black holes (BH), than it was expected 
even a few years ago.}
{They may make a considerable or even 100\%  contribution to the cosmological dark matter.}
{
Among these BH:}
\begin{itemize}
\item
{massive, from a fraction of  ${M_\odot}$ up to ${\geq {  10 M_\odot}}$,}
\item
{supermassive (SMBH), ${ M \sim (10^6 - 10^{9} ) M_\odot }$,}
\item
 intermediate mass (IMBH)
 $ M\sim (10^3 - 10^5) M_\odot $.
\end{itemize}

{Conventional mechanism of creation of these PHs is not efficient.
Most natural is to assume that these black holes are  primordial}, (PBH).
Existence of such abundant primordial black holes was predicted more than a quarter of century ago~\cite{AD-JS}.
Not only abundant PBHs but also peculiar primordial stars, observed now, are predicted.
An extreme claim that (almost) all black holes in the universe are primordial looks quite realistic.

There is large amount of astronomical data, mostly accumulated during several recent years and constantly
appearing almost every day, which are at odds with the accepted standard cosmological model. The review of 
existing state at 2018 is presented in ref.~\cite{AD-UFN}. These data is discussed here together with some more
recent observations.


\section{Types of BH by creation mechanism \label{s-types-BH}}

There are three possible known ways to make a black hole:\\[1mm]
{\it I. Astrophysical BHs: created by stellar collapse when star exhausted its nuclear fuel.} \\
Expected masses 
are just above the neutron star masses ${3 M_\odot}$ and normally they are quite close to it. We observe
instead that  the mass spectrum of BH in the Galaxy has maximum at 
$M \approx 8 M_\odot$   with the width: ${ \sim(1-2) M_\odot }$, see below.
{It is unknown how the traditional mechanism can lead to such surprising 
form of the mass spectrum.}\\[1mm]
{\it II. Accretion of matter to regions with excessive density.} \\
{There are supermassive BHs (SMBH) in all large galaxies with ${M\sim 10^9 M_\odot}$ in elliptic and lenticular 
galaxies and ${M\sim (10^6-10^7) M_\odot} $ in elliptic galaxies, like Milky Way.}
{However, the known mechanisms of accretion are not efficient enough to create such monsters during the universe 
age ${t_U \approx 15 }$ Gyr.}
{Very massive seeds are necessary, but their origin remains mysterious.}

{Moreover SMBH are found in very small galaxies and one SMBH lives even in almost empty space.}
{SMBH are also discovered recently in quite young universe with the age about (1 - 0.5) Gyr.}\\[1mm]
 {\it III. Primordial black holes (PBH) created in pre-stellar epoch, in the very early universe.}\\
{The canonical picture of their formation is the following: 
the density excess might accidentally happen to be large ${\delta\rho/\rho \sim 1}$
at the cosmological horizon scale.} 
{Then this piece happened be inside its gravitational radius i.e. it became a BH,
and decoupled from the cosmological expansion.} 
This mechanism was suggested by Zeldovich and Novikov in 1967~\cite{ZN-BH}, and elaborated later by 
Carr and Hawking in 1974~\cite{CH-BH}.

In traditional approach this mechanism is assumed to create
PBH with rather low masses and with sharp almost delta-function mass 
spectrum. However, cosmological inflation allows for much higher masses and an
extended mass spectrum. In particular, according to the mechanism suggested in ref.~\cite{AD-JS} 
and furfther studied and developed in~ \cite{D-K-K}, 
PBH with masses exceeding millions solar masses with
 a very simple, log-normal mass spectrum, see below eq.~(\ref{dn-dM}),  could be created. 
 
Other early publications on the effects of inflation on PBH creation, resulting in different forms of extended mass spectrum,
include~\cite{I-N-N,GB-L}. They were followed by a  long period of silence  and only recently
a few years ago they attracted the deserved great attention.

\section{Problems in contemporary universe \label{s-today}}

\subsection{Supermassive black holes (SMBH) today \label{ss-smbh-0}}
Every large galaxy and even some much smaller ones 
contains a central supermassive BH with mass oftern larger than 
${ 10^{9}M_\odot}$ in giant elliptical
and compact lenticular galaxies
and {${\sim10^6 M_\odot}$} in spiral galaxies like Milky Way.
The largest mass of BH observed in contemporary universe is 
${M\approx 6\cdot 10^9 M_\odot}$~\cite{smbh-max}.
The origin of these SMBHs is mysterious.

The accepted faith is that
these BHs are created by matter accretion to a central seed of unknown origin.
Moreover, the accretion efficiency is insufficient at least by two orders of magnitude 
to make them during the Universe life-time,  ${t_U = 14.6\cdot 10^9}$.
 
 The accretion efficiency to the central black hole in our Galaxy was calculated  in
 ref.~\cite{murch}. Quoting the authors;
A supermassive black hole SgrA* with the mass ${\sim 4\times10^6 M_\odot}$ resides at the 
centre of our galaxy. Building up such a massive black hole within the ${\sim 10^{10}}$ 
year lifetime of our galaxy would require a mean accretion rate of  ${4\times 10^{-4}} M_\odot$  per year.
At present, X-ray observations constrain the rate of hot gas accretion 
to ${\dot{M} \sim 3 \times 10^{-6} M_\odot}$ per year 
and polarization measurements constrain it near the event horizon to 
${\dot{M}_{horizon} \sim 10^{-8} M_\odot}$/yr. 
The universe age is short at least by two orders of magnitude.

 Even more puzzling is that SMHBs  are observed  in some very small galaxies
 and even in almost {EMPTY} space, where no material to make a SMBH can be found.

An inverted picture of SMBH formation looks more plausible, when first a SMBH was formed and 
attracted matter being a seed for subsequent galaxy formation, as it is suggested in
refs.~\cite{AD-JS,D-K-K,Bosch}.

\subsection{Quasar multiplets \label{ss-multi-QSO}}

Several multiple quasar systems are observed, which have very low probability of formation in the conventional
theory.  \\[1mm]
{\bf Four QSO binaries:} \\
P. Kharb, {\it et al} "A candidate sub-parsec binary black hole in the Seyfert galaxy NGC 7674";
distance d=116 Mpc, mass ${3.63\times 10^7 M_\odot}$~\cite{bin-1};\\
 {C. Rodriguez et al. A compact supermassive binary black hole system. 
 at the distance ${d \approx 230}$~Mpc~\cite{bin-2};}\\
{M.J.Valtonen,"New orbit solutions for the precessing binary black hole model of OJ 287"; redshift: 
${z \approx 0.3}$~\cite{bin-3};}\\ 
{M.J. Graham et al. "A possible close supermassive black-hole binary in a quasar with
optical periodicity";  ${z \approx 0.3}$~\cite{bin-4}.} \\[1mm]
{\bf Triple quasar}:\\ 
E. Kalfountzou,  {\it et al}~\cite{tri}
{" A Triple AGN or an SMBH Recoil Candidate?"} \\
A kiloparsec-scale  supermassive black hole system at z=0.256 is
discovered by systematic search for binary quasars .
The system contains three strong emission-line nuclei, which are offset 
by ${< 250}$ km/s i.e.  by 15-18 kpc in projected separation, 
{suggesting that the nuclei belong to the same physical structure. }
{Quoting  the authors, such a structure can only satisfy one of the three scenarios: a triple supermassive black hole 
interacting system, a triple AGN, or a recoiling SMBH.}\\[1mm]
{\bf Quasar quartet} \\
According to J.F. Hennawi {\it et al}~\cite{quartet}, four quasars, embedded in giant nebula reveal rare massive structure 
in distant universe at  $z\approx 2$.
{The probability of finding a  quadruple quasar is estimated to be 
${\sim 10^{-7} }$.}
 The data imply that the most massive structures in the distant universe have a tremendous supply 
 $\sim 10^{11} M_{\odot}$  of cool dense (volume density $\sim 1$/cm${^3}$) gas, 
 which is in conflict with current cosmological simulations. 

\subsection{ Intermediate mass black holes (MBH)  ${M =(10^3 -10^5) M_\odot}$ \label{ss-inn-M-BH}}

Nobody expected them and now they came out as if from cornucopia (cornu copiae). \\
Four years ago only ten IMBH was known with masses from
$ 3\times10^4 $ up to $2 \times10^5 M_\odot$.
Forty IMBH  with masses $(10^4 - 10^5) M_\odot$
were found in 2018  in dwarf galaxies with stellar masses ${10^7<M<3\cdot 10^9}$~\cite{imbh-dwarf}.
 The same year a sample of  204 IMBHs  in active galactic nuclei was
 presented~\cite{imbh-204}  with black hole masses in the range of
{${(1-20) \times 10^5 M_{\odot}}$.}
 Slightly later 305 IMBH with masses ${{3\times10^4<M_{\mathrm{BH}}<2\times10^5 M_{\odot}}}$
 have been identified~\cite{imbh-305}. 
 A review on IMBH observations is given in refs~\cite{imbh-rev-1,imbh-rev-2}

It is tempting to assume that the intermediate mass PBHs with ${M\sim 10^{4}-10^5}$ are  the 
seeds of dwarf galaxy formation, while less massive ones with ${M \sim 10^3 M_\odot}$, seeded globular clusters.
However, only one or two massive BH are observed in
Globular clusters.
Definite evidence of BH with ${M \approx 2000 M_\odot}$ was found
in the core of the globular cluster 47 Tucanae~\cite{imbh-glob-1} and an evidence for IMBH with $M\sim 10^4 M_\odot$ is 
reported in ref.~\cite{imbh-glob-2}.

The origin of IMBH in the standard model is unknown.
Our prediction~\cite{AD-KP-glob-clrst} is that
if the parameters of the mass distribution of PBHs (see below, eq.~(\ref{dn-dM}))
are chosen to fit the LIGO data and the density of
SMBH, then  the number of PBH with masses 
$(2-3)\times 10^3 M_\odot$ is about ${10^4-10^5 }$ per one SMPBH with mass ${>10^4 M_\odot}$.
This predicted density of IMBHs is sufficient to seed the formation of all globular clusters in galaxies,
as well as the formation of dwarfs.


\subsection{Strange stars \label{ss-strange-stars}}

\subsubsection{Old stars in the Milky Way.}
$ $\\
New more accurate methods of determination of stellar ages led to discovery of
surprisingly old stars, some of them being 
older than the host Galaxy and one star looks even older than the universe.\\ 
Employing thorium and uranium  abundances
in comparison with each other and with several stable elements {the age of
metal-poor, halo star BD+17$^o$ 3248 was estimated as}  { ${13.8\pm 4}$ Gyr}~\cite{old-1}.\\
{For comparison the age of inner halo of the Galaxy} is {${11.4\pm 0.7}$ Gyr}~\cite{old-2}. \\
{The age of a star in the galactic halo, HE 1523-0901, was estimated to be 
about 13.2 Gyr.}
{First time many different chronometers, such as the U/Th, U/Ir, Th/Eu and Th/Os ratios to
measure the star age have been employed~\cite{old-3}}.\\
Metal deficient {high velocity} subgiant in the solar neighborhood
HD 140283  has the age {${14.46 \pm 0.31 }$ Gyr~\cite{old-4}}.
The determined central value of the age exceeds the universe age by two standard deviations 
if the Hubble parameter is low, ${H= 67.3}$ (according to the CMB analysis) and $t_U =13.8$;
{while if ${H= 74}$ (according to the traditional methods), and ${ t_U = 12.5}$, the age of this star 
exceeds the universe age more than by 10 ${\sigma}$.} 

Of course a star cannot be older than the universe. A possible explanation is that according to our 
model~\cite{AD-JS,D-K-K} is that there can be primordial stars enriched with heavy elements, 
so they may look older than they are. 

Another striking example of an unusually old object is is the discovery of a hot rocky planet with
the mass close to that of Neptune with the age: ${10.6^{+1.5}_{-1.3} }$ Gyr~\cite{old-planet}.
For comparison the age  of the Earth is 4.54 Gyr.
A supernovae explosion and molecule and dust formation must precede  formation of this planet.

\subsubsection{Stars with "wrong" chemistry and velocities}
$ $ \\
Several stars with rather unexpected in the conventional astrophysics properties have 
been discovered during last 2-3 years. They have too large velocity, larger than the virial velocity
in the Galaxy, which is about 200  km/sec,
and an unusual chemical  content. There are several very fast pulsars in the Galaxy, but
their origin is evident. Pulsara are the results of supenova explosions and a small angular asymmetry in the
emitted radiation could create a strong kick, which would accelerate a pulsar up to  $~10^3$ km/sec.
The observed fast stars have velocities about 500 km/sec and, otherwise, look normal.

Two years ago a discovery of a low mass white dwarf, LP 40-365,
was reported, which travels at a velocity greater than the Galactic escape velocity
and have peculiar atmosphere which is dominated by intermediate-mass elements~\cite{fast-WD}. 
The origin of this white dwarf is in strong tension with the accepted astrophysics. However, it 
can naturally be a primordial star with high initial abundances of heavy elements~\cite{AD-JS,D-K-K}.

Let us mention several more discoveries of
other high velocity stars in the Galaxy~\cite{fast-2,fast-3}. The authors conclude that  
they can be accelerated by a population of IMBHs in Globular clusters, if there is sufficient number of IMBHs.
So many IMBHs were not expected but the recent data reveal more and more of them in contrast to conventional
expectations and in agreement with ref.~\cite{AD-JS,D-K-K}.

An unusually {\bf red} star was observed in planetary system through microlensing event~\cite{red-star}.  
The host star and planet masses are estimated as 
$M_{\rm host} =0.15^{+0.27}_{-0.10}M_\odot$ and $m_p=18^{+34}_{-12}M_\oplus$.
According to the authors, the life-time of main sequence star with the solar chemical content is larger than 
the universe age already for $ M< 0.8 M_\odot  $. It implies its primordail origin with already evolved chemistry.
May it be a primordial helium  star? There should be stars dominated by helium in our scenario. 

Practically at the date of the conference one more striking discovery was announced!~\cite{fast-WD-2}.
The author's conclusion, is: 
"We report the likely first known example of an unbound white dwarf that is
consistent with being the fully-cooled primary remnant to a Type Iax supernova.
The candidate, LP 93-21, is travelling with a galactocentric velocity of
{${v_{gal} \simeq  605 km \, s^{-1}}$,} and is gravitationally unbound to the Milky
Way, We rule out an extragalactic origin. The Type Iax supernova ejection
scenario is consistent with its peculiar unbound trajectory, given anomalous
elemental abundances are detected in its photosphere via spectroscopic
follow-up. This discovery reflects recent models that suggest stellar ejections
likely occur often."
{However, this event being a remnant of a primordial star is not ruled out.}

\subsection{MACHOs \label{macho}}

MACHOs are invisible (very weakly luminous or even non-luminous) objects {with
masses about a half of the solar mass.} 
They are  discovered through gravitational microlensing  by Macho and Eros groups and 
 later also observed {in the Galactic halo, in the center of the Galaxy, and recently in
the Andromeda (M31) galaxy. }
 

 The observational situation with them is rather controversial and is recently analyzed in
 our paper~\cite{ad-sp}, which we follow here, 
 and in earlier works~\cite{moniez,sib-ufn,Blinnikov:2014nea,bdpp}.
 MACHO group \cite{MACHO2000} reported registration of 13 - 17 microlensing events  towards the Large Magellanic Cloud (LMC),
which is significantly higher than the number which could originate from the known low luminosity stars. On the other hand  this amount
is not sufficient to explain all dark matter in the halo. The fraction of the mass density of the observed objects, which created the
microlensing effects, with respect to the energy density of the dark matter in the galactic halo, $f$, according to the 
observations~\cite{MACHO2000} is in the interval:
\be
0.08<f<0.50 ,
\label{f-macho}
\ee
at 95\% CL for the mass range $  0.15M_\odot < M < 0.9M_\odot  $.


EROS 
collaboration~\cite{EROS-1}  has placed the upper
limit on the halo fraction, $f<0.2$ (95\% CL) for { the} objects in the specified above MACHO
mass range, while EROS-2 \cite{Tisserand:2006zx} gives $ f<0.1$ for $0.6 \times 10^{-7}M_\odot<M<15 M_\odot$
for the survey of Large Magellanic Clouds.
It is considerably less than that measured by the MACHO collaboration in the central region of the LMC.
The data in support  of smaller density of MACHOs in the direction to SMC
is presented in ref.~\cite{Tisserand:2006zx}.

The new  analysis of 2013 
by EROS-2, OGLE-II, and OGLE-III collaborations~\cite{Novati:2013fxa} towards the Small Magellanic Cloud (SMC)
revealed five  microlensing events towards the SMC (one by EROS and four by OGLE), which lead to the upper limits
 $ {f <0.1} $ obtained at 95\% confidence level for MACHO's with the mass $ 10^{-2} M_\odot$
and $ {f <0.2} $ for MACHOs with the mass $ 0.5 M_\odot$.

Search for microlensing in the direction of Andromeda galaxy (M31) demonstrated some contradicting 
results~\cite{moniez,sib-ufn} with an uncertain conclusion. E.g. AGAPE collaboration \cite{AGAPE2008}, 
finds the halo MACHO fraction in the range $0.2<f<0.9$.
while MEGA group presented the upper limit $f<0.3$~\cite{MEGA2007}.
On the other hand, the recent discovery of  10 new microlensing events~\cite{Lee-2015}  
is very much in favor of MACHO existence. The authors conclude: ``statistical studies and individual microlensing events
point to a non-negligible MACHO population, though the fraction in the halo mass remains uncertain''. \\
Some more recent observational data and  the other aspects  of the microlensing are discussed in ref.~\cite{Mao2012}.
\\[1mm]
 To summarize:\\
Macho group:
 $ 0.08<f<0.50 $ (95\% CL)  {for $ 0.15M_\odot < M < 0.9M_\odot $;}\\
EROS: {${f<0.2}$, ${ 0.15M_\odot < M < 0.9M_\odot}$;}\\
EROS2:{ $ { f<0.1}$,~${10^{-6}M_\odot<M<M_\odot}$;}\\
AGAPE: {${0.2<f<0.9}$ for ${0.15M_\odot < M < 0.9M_\odot} $;}\\
EROS-2 and OGLE: {$ {f <0.1} $ for  ${M\sim 10^{-2} M_\odot}$ and
${f <0.2} $ for  ${ \sim 0.5 M_\odot}$.}\\
MACHOs surely exist, but their density, is not well known. 
Anyhow their density is  significantly greater than the density expected from the known low luminosity
stars and the expected density of BH of similar mass. But PBH may have similar, though not well known
density.

Our attempts~\cite{ad-sp} to obtain the  number density of MACHOs using the log-normal mass spectrum and
adjusting its parameters from different pieces of other data, such as the number density of SMBH, mass spectrum 
of black holes in the Galaxy (see the next section), etc, always led to a very low density of MACHOs. Possible resolutions
to this conundrum is either a superposition of several log-normal spectra with different  values maxima~(\ref{multi-log})
or possible clusterization of MACHOs. The latter assumption explains inconsistency of MACHO observation or 
non-observation of them in different directions to the sky.

\subsection{Mass spectrum of astrophysical (?) BHs in the Galaxy \label{mass-spetrum-gal}}

As it is calculated in ref.~\cite{BH-gal-mass}, theoretically expected mass distribution of galactic black holes has maximum
at the minimal possible value of BH mass at $3 M_\odot$ and exponentially drops down with increasing mass.
According to the authors, no evidence for a gap at low values $(3-5) M_\odot$ or for a peak at higher 
$\sim 7 M_\odot$ is found. 

These theoretical results are in strong conflict with observations. As is stated in paper~\cite{BH-mass-obs-1}
six of the seven systems with measured mass functions have black hole masses
clustered near seven solar masses. There appears to be a significant gap between
the masses of these systems and those of the observed neutron stars.

This result is strongly confirmed by  subsequent observations. It was found~\cite{BH-mass-obs-2}
that the BH masses are concentrated in the narrow range ${ (7.8 \pm 1.2) M_\odot }$
in very good agreement with another paper~\cite{BH-mass-obs-3}
 where {a peak around ${8M_\odot}$, a paucity of sources with masses below
 ${5M_\odot}$, and a sharp drop-off above
${10M_\odot}$ are observed}.\\
{These features are not easily explained  in the standard model of BH
formation by stellar collapse,} {but excellently fit the hypothesis of their primordial origin.}

\subsection{Gravitational waves from BH binaries \label{s-GW-BH-bin}}

Registration of gravitational waves (GW) from BH binaries by LIGO revealed several problems,
which are most naturally solved if one assumes that the sources of GW are primordial black
holes, see e.g.~\cite{bdpp}, as well as a lot of other publications.

\subsubsection{Origin of massive BH,  ${M \sim 30 M_\odot}$ \label{sss-orogin-BH}}
 $ $ \\
 Such BHs, if they are astrophysical, are believed
to be created by massive star collapse, though a convincing conventional theory is still lacking.
{To form so heavy BHs, the progenitors should have huge mass, ${M > 100 M_\odot}$
and  a low metal abundance to avoid too much
mass loss during the evolution.} Such heavy stars might be present in
young star-forming galaxies {but they are not observed in the necessary amount.}
{Primordial BH with the observed by LIGO masses may be easily created with sufficient density. 

The problem of astrophysical BH formation becomes multifold more severe if the black hole with 
mass $M= 68^{+ 11}_{-13}\, M_\odot$ is indeed discovered~\cite{mass70}.

 \subsubsection{Formation of BH binaries from the original stellar binaries. \label{sss-bin-form}}
 $ $ \\
Stellar binaries are supposed to be
formed from a common interstellar gas cloud and are quite frequent in galaxies.
{If BH is created through stellar collapse,} {a small non-sphericity results in a huge 
velocity of the BH and the binary would be destroyed.} 
{BH formation from PopIII stars and subsequent formation of BH
binaries with 
${(36+29) M_\odot}$ is analyzed and 
found to be negligible. } \\
{The problem of the binary formation is simply solved if the observed sources of GWs are the binaries of
primordial black holes.} 
They were at rest in the comoving volume and when inside horizon they are gravitationally attracted 
and may loose energy due to dynamical friction in the early universe.
The probability of mutual capture and forming binaries of PBHs may be large enough.

\subsubsection{Low spins of the coalescing BHs \label{sss-spin}}
$ $\\
The low values of the BH spins in GW150914 and in almost all, except for 2-3, other events,
strongly constrain astrophysical BH formation from close binary systems. 
{Astrophysical BHs are expected to have considerable angular momentum, nevertheless the
dynamical formation of double massive low-spin BHs in dense stellar clusters is not excluded, though difficult.} 
{On the other hand, PBH practically do  not rotate because vorticity perturbations 
in the early universe are vanishingly small.}\\
{However, individual PBH forming a binary initially rotating on elliptic orbit could gain collinear spins about 0.1 - 0.3,
rising with the PBH masses and eccentricity~\cite{pm-spin-1,pm-spin-2}
{This result is in agreement with the
GW170729 LIGO event produced by the binary with masses ${50 M_\odot}$  and ${30 M_\odot}$ 
and  maybe with GW151216.}

In earlier works~\cite{low-spin-1,low-spin-2} much weaker gain of angular momentum was claimed.

\section{ Young universe, ${\bf z=5-10}$. A brief review of high-z surprises. \label{s-young}}

\subsection{Early galaxies \label{ss-earluy-gal}}

Several galaxies have been observed at high redshifts,
with natural gravitational lens ``telescopes",
for example a galaxy at ${z \approx 9.6}$ which was created 
when the universe age was about   $t_U \approx \,0.5$ Gyr~\cite{gal-9.6}. \\
{A galaxy at {${z \approx 11}$} has been detected ~\cite{gal-11}
which was formed earlier than the universe reached  ${ 0.41}$ Gyr. 
(or even shorter with large H).
This galaxy is three times more luminous in UV than other galaxies at ${z = 6-8}$.} 
 It is surprising that so bright galaxy was created in so short time.\\

Not so young, at  ${t_U \sim 1.3 }$ Gyr, but extremely luminous galaxy was found~\cite{gal-max-lum}
with the luminosity {${L= 3\cdot 10^{14} L_\odot }$.
{For its creation galactic seeds, or embryonic black holes, might be bigger than thought possible.}
Quoting one of the authors, P. Eisenhardt: "How do you get an elephant?  One way is start with a baby elephant."
However, there is no known mechanism in the standard model to make sufficiently heavy seeds.
{The mass of the BH seed  should be already billions of ${M_\odot}$ , when our universe was only a 
tenth of its present age of 13.8 billion years.}
As mentioned in the paper, another way to grow this big is to have gone on a sustained binge, consuming food 
faster than typically thought possible.  
The necessary condition for the fast rise of the mass is a low spin of  the BH. As is mentioned in 
subsubsection \ref{sss-spin}, low spin is a strong indication that the black hole is primordial.

A large population of massive galaxies in the early universe  at $z > 3$ is described 
in ref.~\cite{wang-gal}. The detection of 39 massive star-forming galaxies in submillimeter range
(wavelength 870{$\mu$}m) is reported there. Quoting the paper,
these galaxies are unseen in the spectral
region from the deepest ultraviolet to the near-infrared. 
They contribute a total star-formation-rate density ten times larger than that of equivalently massive
ultraviolet-bright galaxies at $z >3$. Residing in the most massive dark matter
halos at their redshifts, they are probably the progenitors of the largest
present-day galaxies in massive groups and clusters. 

{Such a high abundance of
massive and dusty galaxies in the early universe challenges our understanding
of massive-galaxy formation.}

As is stated in the paper "Monsters in the Dark"~\cite{monsters}, 
density of galaxies at ${z \approx 11}$ is 
$10^{-6} $ Mpc${^{-3}}$, an order of magnitude higher than estimated from the data at lower z.
{Origin of these galaxies is unclear.}

According to F. Melia~\cite{melia} 
{"Rapid emergence of high-z galaxies so soon after big bang} 
may actually be in conflict with current understanding of how they came to be. This
problem is very reminiscent of the better known (and
probably related) premature appearance of supermassive
black holes at $z\sim 6$. It is difficult to understand how
{$10^9 M_\odot$} black holes appeared so quickly after the big
bang {without invoking non-standard accretion physics and the formation of massive seeds, 
both of which are not seen in the local Universe."

\subsection{ Supermassive BH and/or QSO \label{ss-BH-QSO}}

{Another and even more striking example of early formed objects are high z quasars.}
About 40 quasars with ${z> 6}$ were known four years ago, each quasar containing BH with 
${M \sim 10^9 M_\odot}$. 
The maximum quasar redshift  ${ z = 7.085}$ QSO is discovered in ref.~\cite{QSO-7085} 
with {${L \approx 6 \cdot 10^{13} L_\odot}$} and {${M=2 \cdot 10^9 M_\odot}$,} 
The quasar was formed before the universe came to the age {${0.75}$ Gyr. } 

In addition to all that another monster was discovered at redshift 6.30 and mass twelve billion solar mass~\cite{12-QSO}. 
{There is already a serious problem with formation of lighter and less luminous quasars}
{which is multifold deepened with this new "creature".}
This huge mass  ${M \approx 10^{10} M_\odot }$  makes the formation
{absolutely impossible in the standard approach by accretion to some matter excess.} 

Moreover, as follows for the results of the paper~\cite{neutral}: 
{"An 800 million solar mass black hole in a significantly neutral universe at redshift 7.5",} 
 Any significant accretion leads to ionization of the surrounding matter. Matter neutrality means that
 accretion is practically absent


The accretion rate in the early universe was calculated by Latif, Volonteri, and Wise~\cite{accr-rate}, 
who have found that
''.. halo has a mass of ${ 3 \times 10^{10}~M_{\odot}}$ at ${z=7.5}$;
MBH accretes only about 2200 ${M_{\odot}}$ during 320 Myr", which is by far below the necessary amount.

To conclude on QSO/SMBH:
{the formation of quasars, or what is the same, of supermassive black holes,}
{ in such short time by conventional mechanisms looks problematic to say the least.} {Such black holes,
{when the Universe was less than one billion years old,} 
present substantial challenges to theories of the formation and growth of
black holes and the coevolution of black holes and galaxies.}
{Even the origin of SMBH in contemporary universe during much longer time $t_U= 14$ Gyr is unclear.} 



\subsection{Evolved chemistry, dust, supernovae, gamma-bursters, {\it etc} \label{s-dust}}

{The medium around the observed early quasars contains
considerable amount of ``metals''} (elements heavier than He). 
According to the standard picture, only elements up to ${^4}$He  { and traces of Li, Be, B}
were formed by BBN, {while heavier elements were created
by stellar nucleosynthesis and} {dispersed in the interstellar space by supernova explosions.}
{Hence, an evident but not necessarily true conclusion was
that prior to or simultaneously with the QSO formation a rapid star formation should take place.}
{These stars should evolve to a large number of
supernovae enriching interstellar space by metals through their explosions.}

{Another possibility is a non-standard BBN in bubbles with very high baryonic density~\cite{AD-JS,D-K-K}, 
which allows for primordial formation of heavy elements beyond lithium.}

According to numerous recent observations the universe at ${z >6}$ is unexpectedly full of dust~\cite{mancini}. 
Abundant dust is observed also in the  observations~\cite{Mattsson,Venemans}.
{Dusty galaxies show up at redshifts corresponding
to a Universe which is only about 500 Myr old.}
Several early  galaxies, e.g. in HFLS3 at ${ z=6.34} $ and in
A1689-zD1 at ${ z = 7.55}$ are also full of dust.
Past high star formation is needed to explain the presence of $ \sim10^8 M_\odot$ of dust implied by the 
observations~\cite{Venemans}.

The amount of the  observed dusty sources is an order of magnitude larger 
than  that predicted by the canonical theory.

{To make dust a long succession of processes is necessary:} 
{first, supernovae  explode to deliver 
heavy elements into space (metals),} 
{then metals cool and form molecules,}
{and lastly molecules make dust which could form macroscopic
pieces of matter,}  turning subsequently into early rocky planets.\\
{We all are dust from SN explosions, at much later time} 
{but abundant dust may indicate {\bf that there also could be life 
in the early.} Several hundred million years may be enough for creation of living creatures. }

The summary of dust production scenarios at high redshifts,  ${z \sim  6-8.3}$  presented in~\cite{dust-sum}
is the following.
 {The mechanism of dust formation in galaxies at high redshift is still
unknown.} Asymptotic giant branch (AGB) stars and explosions of supernovae (SNe)
are possible dust producers, and non-stellar processes may substantially
contribute to dust production. However, AGB are not efficient enough
to produce the amounts of dust observed in the galaxies. 
{In order to explain these dust masses, SNe would have to have maximum efficiency and not
destroy the dust which they formed.} Therefore, the observed amounts of dust in
the galaxies in the early universe were formed either by efficient supernovae
{or by a non-stellar mechanism, for instance the grain growth in the
interstellar medium.}

{Another option is the  non-standard big bang nucleosynthesis with large baryon-to-$\gamma$ ratio leading to 
abundant formation of heavy elements.}

{Observations of high redshift gamma ray bursters (GBR) also indicate 
a high abundance of supernova at large redshifts.} 
{The highest redshift of the observed GBR is 9.4 and there are a few more
GBRs with smaller but still high redshifts.}
{The necessary star formation rate for explanation of these early
GBRs is at odds with the canonical star formation theory.}

Again the non-standard big bang nucleosynthesis with large baryon-to-photon ratio leading to formation of 
heavy elements may easily help.

\section{Creation mechanism \label{s-create}}

The mechanism of massive PBH formation with wide mass spectrum was proposed and developed 
in refs~\cite{AD-JS} and \cite{D-K-K} respectively
 Heretic predictions of 1993} 
{are turning now into the accepted faith, since they became
supported by  astronomical data.}
 Massive PBHs allow to cure emerging inconsistencies with 
the standard cosmology and astrophysics.}
Dark matter made out of PBHs became a viable option.
{The model  predicts an abundant formation of heavy PBHs with log-normal mass spectrum:} 
\be
{dN}/{dM} = \mu^2 \exp{[-\gamma \ln^2 (M/M_0)], }
\label{dn-dM}
\ee
with only 3 parameters: ${\mu}$, ${\gamma}$,  ${M_0}$. 
{The spectrum can be generalized to multi-maximum spectrum i.e. to superposition of log-normal spectra with 
different $M_0$.}
{Log-normal spectrum is a result of  quantum 
diffusion of baryonic scalar field during inflation. Probably such spectrum is a general consequence of diffusion.}

The concrete calculations are based on the so called  supersymmetry (SUSY) motivated or Affleck and Dine (AD)
 baryogenesis~\cite{AD}.
SUSY predicts existence of  scalar bosons, $\chi$, with non-zero baryonic number,
{${ B\neq 0}$.} The potential of these bosons generically has flat directions, along which it does not rise:
\be
U_\lambda(\chi) = \lambda |\chi|^4 \left( 1- \cos 4\theta \right) .
\ee

There can be also the mass term, ${ m^2 \chi^2 + m^{*\,2}\chi^{*\,2}}$:
\be
U_m( \chi ) = m^2 |\chi|^2} {\left[{ 1-\cos (2\theta+2\alpha)} 
\right],
\nonumber
\ee
where ${ \chi = |\chi| \exp (i\theta)}$ and ${ m=|m|e^\alpha}$. {If ${\alpha \neq 0}$, C and CP are  broken.}

The field $\chi$ may condense along {flat} directions of the quartic potential, at the stage when the Hubble parameter
is much larger than its mass.

{In grand unified SUSY models baryonic number is naturally non-conserved, which is reflected  in non-invariance 
of ${U(\chi)}$ w.r.t. phase rotation, $\chi \rar \chi \exp (i \omega)$ with a constant phase $\omega$.


{ Initially (after inflation) ${\chi}$ is away from origin and, when 
inflation is over, starts to evolve down to equilibrium point, ${\chi =0}$,
according to equation  similar to that in Newtonian mechanics:}
\be
\ddot \chi +3H\dot \chi +U' (\chi) = 0.
\ee
Baryonic charge of $\chi$ is defined as:
\be
B_\chi =\dot\theta |\chi|^2
\ee
It is analogous to mechanical angular momentum in potential $U(\chi)$. If in the process of cosmological evolution
field $\chi$ started to rotate in complex $\chi$ plane. It means that $\chi$ acquired baryonic number, which is usually large.
 The decay of ${{\chi}}$ transferred the accumulated baryonic number to that of quarks in B-conserving process.
Correspondingly AD baryogenesis could lead to baryon asymmetry of order of unity, much larger
than the observed  ${\beta = n_B/n_\gamma =6\times 10^{-10}}$.

If $ { m\neq 0}$, the angular momentum, or B, is generated due to different 
directions of the  quartic and quadratic valleys. Moving along a quartic valley at high $\chi$ down to low $\chi$
the field started to "feel" quadratic valley and begun attracted  towards it. That's how rotation or $B$  can be 
generated

{If CP-odd phase ${\alpha}$ is small but non-vanishing, both baryonic and 
antibaryonic domains might be  formed}
{with possible dominance of one of them.}
Matter and antimatter domains may exist but globally ${ B\neq 0}$.
 
We have modified the AD baryogenesis adding general renormalizable coupling of field $\chi$ to the inflaton $\Phi$,
the first term in the equation below:
\be 
U = {g|\chi|^2 (\Phi -\Phi_1)^2}  +
\lambda |\chi|^4 \,\ln ( \frac{|\chi|^2 }{\sigma^2 } ),
{+\lambda_1 (\chi^4 + h.c. ) + 
(m^2 \chi^2 + h.c.). \,\,\,\,\,\,\,\,\,\,\,\,\,\,\,\,\,\,\,\,
\label{U-of-chi-Phi}
}\ee
where $\Phi_1$ is the value of $\Phi$, which it passed during inflation,  and the second term is a result of one-loop 
corrections to the original bare potential, the Coleman-Weinberg correction.
CP would be broken, if the relative phase of ${\lambda_1}$ and 
$m$ is non-zero, otherwise one can
``phase rotate'' $\chi$ and come to real coefficients. 
{Coupling of $\chi$  to fermions may also break CP.}

{When the window to the flat direction is open, near ${\Phi = \Phi_1}$, }
{the field ${\chi}$ started to diffuse to large value,} according to quantum diffusion
equation derived by Starobinsky~\cite{diff-AAS-1,diff-AAS-2}, generalized 
in our works to a complex field ${\chi}$.

If the window to flat direction, when ${\Phi \approx \Phi_1}$ is open only {during 
a short period,} cosmologically small but possibly astronomically large 
bubbles with high ${ \beta}$ could be
created, occupying {a minor
fraction of the universe volume,} while the rest of the universe  would have the normal
{${{ \beta \approx 6\cdot 10^{-10}}}$, created 
by small ${\chi}$}. 

{This mechanism of massive PBH formation is quite different from previously studied  ones.}
{The fundament of PBH creation was build at inflation by making large isocurvature
fluctuations at relatively small scales, with practically vanishing density perturbations.} 

The initial isocurvature perturbations are contained in density contrast of massless quarks and antiquarks.
Density perturbations arose rather late after the QCD phase transition at temperatures $T\sim 100$ MeV.
{The emerging universe looks like a piece of Swiss cheese, where holes are high baryonic 
density objects occupying a very small fraction of the universe volume.}\\


{The outcome of this process, depending on ${\beta = n_B/n_\gamma}$, is the following:}
\begin{itemize}
\item
PBHs with log-normal mass spectrum. The following  modification of $\chi$ interaction with the 
inflaton:
\be 
U = |\chi|^2 \sum_j^N \lambda_j \Pi_j ^N(\Phi - \Phi_j)^2 
\label{multi-log}
\ee
would create a superposition of $N$ log-normal mass spectra with different maxima.

\item
{Compact stellar-like objects, similar e.g. to the  cores of red giants.}
\item
{Disperse hydrogen and helium clouds  with (much) higher than average ${n_B}$ density.}
\item
{${\beta}$ may be negative leading to compact antistars which could survive annihilation with the 
homogeneous baryonic background.}
\end{itemize}



\section{Conclusion \label{s-concl}}

We predicted or explained the following pieces of data or phenomena:
\begin{itemize}
\item
{ 1. 
 Abundant formation of PBHs and compact stellar-like
objects in the early universe after QCD phase transition, ${t \geq 10^{-5}} $ sec.}
\item
{ 2. Log-normal mass spectrum of these objects.} 
\item
 3.  The peculiar features of the sources of GWs observed by LIGO.
\item
 4. Solution of the numerous mysteries of ${z \sim 10}$ universe: abundant population of supermassive black holes, 
gamma-bursters, supernovae, and early bright galaxies, as well as evolved chemistry including dust.
\item
{5. Suggestion of the inverted picture of galaxy formation, when first a supermassive 
BH seeds  were created and later they accreted  matter forming galaxies.}
\item
{6. An explanation of existence of supermassive black holes observed  arge and some small galaxies and even in
almost empty environment.}
\item
{7. Mechanism of formation of stars older than the universe. }
\item
{8. Existence MACHOs.} 
\item
{ 9. An explanation of origin of BHs with 2000 ${M_\odot}$ in the cores of globular clusters and the
observed density of global clusters.}
\item
{10. Prediction of a large number of  IMBHs.}
\item
{11. Suggestion that dark matter can consist of PBHs.}
\item
{12. A possible existence of abundant antimatter in the Galaxy,}
\end{itemize}



\section*{References}

\section*{Acknowledgement} 
The work  was supported by the RSF Grant 19-42-02004. 

\end{document}